\documentclass[aps,prl,twocolumn,10 pt,showpacs]{revtex4}
\usepackage{amssymb}
\usepackage{graphicx}

\begin{document}

\title{transport and magnetic properties in $YBaCo_2O_{5.45}$: focus on the high temperature transition.}
\author{A.Pautrat, P. Boullay, S. H$\acute{e}$bert, V. Caignaert}
\affiliation{Laboratoire CRISMAT, UMR 6508 du CNRS et de l'Ensicaen, 6 Bd Marechal Juin, 14050 Caen, France.}

\begin{abstract}

The electronic transport properties and the magnetic susceptibility were measured in detail in $YBaCo_2O_{5.45}$.
Close to the so-called metal-insulator transition, strong effects of resistance relaxation, a clear
thermal hysteresis and a sudden increase of the resistance noise are observed. This is likely due to the first order
 character of the transition and to the underlying phases coexistence. Despite these out of equilibrium features, a positive and linear magneto-resistance is also observed,
 possibly linked to the heterogeneity of the state. From a magnetic point of view, the paramagnetic to ordered magnetic state transition is observed using non linear susceptibilty.
This transition shows the characteristics of a continuous transition, and time dependent effects can be linked with the dynamics of magnetic domains in presence of disorder.
 Thus, when focusing on the order of the transitions, the electronic one and the magnetic one can not be directly associated.

\end{abstract}
\newpage
\maketitle

\section{introduction}
The $LnBaCo_2O_{5+x}$ ($Ln$=rare earth) family presents intriguing physical properties such as charge-ordering
 \cite{vogt,suard},
metal-insulator transition \cite{troy}, magnetoresistance \cite{maignan2}.
 One reason to explain this richness of properties is given by the various valences, spin states and coordination of cobalt ions.
 Depending on $x$, Co coexists in octahedral and pyramidal environments which may favour different spin states for a fixed Co valency.
 Focusing on the ideal $x=$ 0.5 stoichiometry, only Co$^{3+}$ should exist. Nevertheless, it is always difficult to neglect some oxygen
 inhomogeneity \cite{plakt}. Numerous reports have already been
 published on the magnetic and transport properties in these compounds.
 The macroscopic magnetic measurements show successive transitions close to room temperature: a paramagnetic to "ferromagnetic" (PM-FM)
 at high temperature, $T_1$, and a "ferromagnetic" to antiferromagnetic (FM-AFM) at lower temperature, $T_2$ \cite{maignan2}.
 Close to $T_1$, a "metal to insulator" transition (M-I) is also observed \cite{maignan2}.
The ground state of the magnetic phases, but also the genuine nature of the "M-I" transition are still a matter of debate.
 In the "ferromagnetic" phase between $T_1$ and $T_2$, the magnetization is very small
 and this could correspond to some Canted Antiferromagnetism \cite{akahoshi}. A detailed single crystal study of $GdBaCo_2O_{5.5}$ has revealed the existence of a structural
 transition at $T_1$, thanks to the observation of low intensity superstructure reflections \cite{chernen},
 and a ferrimagnetic structure is proposed below $T_1$ for $TbBaCo_2O_{5.5}$.
 A spin state transition at $T_1$ between
 a High Spin Co$^{3+}$ and an intermediate Spin Co$^{3+}$ can be proposed.
Indeed, susceptibility measurements in $TbBaCo_2O_{5.5}$ show a change of the Curie constant,
 which supports an explanation in terms of this spin state transition \cite{morito}.
 In $HoBaCo_2O_{5.5}$, the M-I transition has been also analyzed in term of such a spin state transition responsible
 for a 'spin blockade' mechanism \cite{maignan}. 
 In contrast to the interpretation in terms of spin
 state transition, oxygen isotope effect
suggests delocalisation of holes as a possible mechanism which drives the "M-I" transition \cite{delocalisation}, in agreement 
with density-functional calculations \cite{wu}.
When dealing with spin state transition of the Co ions, the interpretation is always difficult when the A site cation induces by itself a magnetic contribution.
 By studying the $YBaCo_2O_{5+x}$ family, this contribution is suppressed and the magnetism should come from the Co ions only.
 We have therefore decided to investigate the magnetic and electronic transport properties of $YBaCo_2O_{5+x}$.  
By focusing on transport properties close to $T_1$, we observe strong out of equilibrium features. In particular,
 long relaxation times and two states-like resistance noise are shown and discussed as features of the first order character of the M-I transition.
 When dealing with quasi-static values of the resistance,
 a previously unreported positive magnetoresistance is observed, which is interpreted as coming from the heterogeneity of the state. 
Ac susceptibility evidences a frequency dependence typical of domains wall pinning, underlying a strong role of disorder and of collective effects.
Finally, despite the fact that both magnetic and transport properties exhibit out of equilibrium features, these latter are notably different. 
We conclude that no direct link can be made between these electronic
 and magnetic transitions, at far as the order of the transition and the states kinetics are involved.

\section{Experimental}
$YBaCo_2O_{5+x}$ was prepared by solid-state reaction. A stoichiometric mixture of $Y_2O_3$, $Co_3O_4$ and $BaCO_3$
 was reacted in air at temperatures from $900$ to $1150^{\circ}$ C with several intermediate grindings.
 After firing at $1150^{\circ}$ C, the sample was slowly cooled to room temperature.
The sample synthesized in air was then fired in oxygen under 10 bars at $350^{\circ}$ C for 12 hours, followed by slow cooling to room temperature.
 The final oxygen stoichiometry was found to be $O_{5.45}$ ($\pm 0.03$) by iodometric titration.
The resistance was measured with the four-probe technique on a $2 \times 2 \times 10$ mm$^3$ bar,
 with a home-made set-up using a PPMS Quantum Design as a cryostat.
 AC and DC magnetic properties were measured using a Squid based magnetometer (MPMS Quantum Design).
The transmission electron microscopy (TEM) study was carried out using a JEOL 2010F electron microscope.
 For TEM observations, fragments of the bulk sample were crushed in an agate mortar and put in a suspension
 in alcohol. A drop of the suspension was then deposited and dried on a copper grid previously coated with a thin film of amorphous carbon. 
Selected Area Electron Diffraction (SAED) investigations show a set of strong reflections corresponding to a 
$a_p$ x $a_p$ x $2a_p$ cell confirming that the sample is a A-site ordered 112-type structure. 
In addition, two distinct sets of weaker reflections can be evidenced depending on the crystal and corresponding 
to a $3a_p$ x $3a_p$ x $2a_p$ and a $2a_p$ x $a_p$ x $2a_p$ supercell (Fig.1). 
As discussed in \cite{akahoshi}, these two superstructures are related to a oxygen/vacancy
ordering and correspond, respectively, to the stoechiometric compositions YBaCo$_2$O$_{5.44}$ and YBaCo$_2$O$_{5.50}$
 (in this latter paper, the oxygen content has been determined by thermogravimetric $H_2$ reduction analysis (TGA)). 
Taken into account the oxygen content that we have measured by iodometric titration, and even through TGA and iodometric
 titration can give slighty different oxygen contents in the case of Cobalt oxides \cite{karpinen}, the SAED patterns shows that the sample under 
investigation could be in the oxygen content range [5.44-5.48] where these two types of superstructure can coexist.  
ED diffraction is in progress to understand if a link between the physical properties and the thermal behavior of this local microstructure can be evidenced.

\section {transport properties and first order transition}

 In fig.2a is shown the variation of the resistance $R$ as a function of the temperature.
 At $T_1\approx $ 290 $K$, the "M-I" is observed,
 as previously reported in other isostructural compounds \cite{maignan2}. 
 We have also plotted $R$ as a function of $T^{-1/4}$ in a semi-log plot in fig.2b. At high and at low temperature,
 one can see a linear slope which is typical of a 3D variable range hopping (VRH), as it
 was observed below the Neel temperature in $GdBaCo_2O_{5.5}$ \cite{raquet}.
 It is worth noting that the M-I transition looks more like
 a cross-over between two semiconducting regimes, than like a genuine M-I transition . In particular, it seems difficult
 to call the regime at $T \geq T_1$ metallic, because $\rho$ is large ($\approx$ 30
$m\Omega$.cm at 300 $K$) and $d\rho/dT$
 is negative, and not positive as it should be for a metal. A VRH or a polaronic behavior, as proposed in \cite{polaron}, are equally compatible with
 our data for the restricted high temperature range which we have meausred.

 We will focus on the nature of the high temperature transition at $T_1$. 
In fig.3, the resistance is plotted as a function of $T$ close to $T_{1}$. 
With a conventional cooling/heating rate (about 1 $K/min$), a large hysteresis is observed, as already reported \cite{maignan2}.
Even through it is reduced, a robust hysteresis is always observed after a very slow cooling/heating rate ($<$ 0.05 $K/min$), 
showing that it does not come
 from training effects of pure thermal origin. An abrupt change of a physical property coupled
 with a thermal hysteresis suggest a first order transition.
 In such a case, the hysteresis comes from the existence of supercooled and superheated states.
The high-temperature phase remains metastable
down to the supercooling temperature and the low temperature phase remains
metastable up to the superheating temperature. Note that, here, the transition is not extremely sharp and extends over several degrees,
 what can question its genuine first order character. However, in real systems, the frozen random disorder can
 not be neglected (cationic and oxygen disorder for example), and the transition can be softened, even if still retaining
 the first order-like characteristics \cite{imry,timonin}.
 
Supercooled states and superheated states are not equilibrium states. Consequently, relaxation effects
 due to the temperature dependent barriers can be expected. Note that, to ensure thermal equilibrium, a waiting time of at least 15 minutes was necessary even after
 the apparent stabilization of the temperature of the PPMS. The reason is the strong 
coefficient $dR/dT$ close to the transition which make the measurement very sensitive
 to an even weak veriation of temperature. This abnormal relaxation can be clearly identified
at the beginning of each measurement by looking at the time series of resistance $R(t)$.
Apart from this spurious effect, an intrinsic relaxation of the resistance is clearly observed.
 The direction of this relaxation depends on the thermal history (not shown).
After cooling the sample, the resistance slowly goes up over a long time scale 
(at least 1 Hour). On the contrary, it slowly goes down after heating the sample.
 For the supercooled states, a good fitting of the data can be obtained with a model of limited growth,
 using a functional form
($R(t)$=$R_0$ (1-$A$.$exp$(-$t$/ $\tau$ ))) where $R_0$ and $A$ are constants (see fig.3). Such a behavior is consistent
 with the evolution of a metastable state into equilibrium.
The parameter $\tau^{-1}$ can be taken as the typical growth rate, and can be extracted by fitting the curves.
Its thermal variation is shown in fig.5. It defines the domain where metastability exists
 after a cooling of the sample, and it can be called a spinodal domain for the supercooled state \cite{spinodal}.
 
 In addition to these relaxation effects, we observe that resistance noise appears when the temperature decreases. 
If the temperature is less than $T$ $\leq $ 290 $K$, some structures can be observed on the $R(t)$ curves.
 They appear more clearly after a subtraction of the main value of $R(t)$ (see Fig.6). 
 Even if these features are not genuine switches between two states (like a telegraph noise for example),
 a strong tendency to a bimodal and non Gaussian behavior can be observed.
 The noise is maximum close to the M-I transition. Such noisy features could originate
 from the fluctuations of magnetic domains/clusters
 close to the paramagnetic to ferromagnetic transition.
 However, we observe that the noise does not change significantly after that a 7 $T$ magnetic field is applied.
 The fact that the noise does not decrease with the magnetic polarization is a strong indication of a non magnetic origin. 
Another possibility is that this shows a phase coexistence due to the first order transition. In such a case, their respective quantities
 are fixed by the thermal history, one can expect hysteresis effects on their kinetics.
We observe indeed that the noise is clearly different after that the sample is (super)cooled or (super)heated.
For example, at $T=$ 289 $K$, a large noise with
 quasi-bimodal histogram is observed for the first case, whereas Gaussian
 and moderate noise is observed for the latter case (not shown). This makes sense because at this temperature,
 this second state is moderately superheated and involve few heterogeneities, contrary to the first case. This thermal dependence
 shows that the noise comes from a mixture of two phases, and not just from sample inhomogeneities.
 The magnitude of the resistance fluctuations $\Delta R$, compared to the main value of $R$, is very large ($\Delta R /R \leq $ 3.10$^{-4}$),
 showing that the current should be highly inhomogeneous in the transition region \cite{mike}. All this consolidates the idea that
 the macroscopic state is heterogeneous, with a coexistence of two zones
with different conductivity.

  For $T <$ 280 $K$, a negative magnetoresistance (MR) appears, being large at low temperature, as already
 reported in others 112 isostructural compounds \cite{maignan,raquet}.
 Not reported before is the moderate, positive and quasi-linear MR that we observe
 close to the M-I transition, i.e. near room temperature. Note that the effect
 of relaxation, which was described before, exist also during the magnetic field cycling.
 This leads to hysteresis properties because the resistance values 
 are changing with time.
We claim that there is a significant equilibrium MR because it is unambiguous higher than this time-dependent part of the hysteresis.
 This MR is positive, moderate and appears only in a restricted temperature range (fig.8). Its magnitude is about (0.5 $^{\circ}/_{\circ}$ /T). This is a reasonable value for conventional galvanomagnetic effects.
 Nevertheless, the $B^2$ dependence typical of the Lorentz force driven mechanism is not observed. Interestingly, there is no clear sign of saturation up to 14 $T$.
 One model of linear MR is the Abrikosov model of "Quantum resistance" \cite{abriko}. Quantum interferences have been also proposed for explaining a positive MR.
 However, it is clear that such analysis are hardly applicable close to room temperature in a ceramic \cite{lmr}.
 Grain boundaries scattering and disorder effects can be significant in our sample.
 Generally, the most simple effect of
grain boundaries is a decrease of the electronic mean free path, leading to a decrease of MR effects in conventional metals.
 However, some experiments point also the role of grain boundaries for the positive MR \cite{gb}. 
A linear MR which does not saturate at high field has been observed in doped silver chalcogenides \cite{lmr}.
 It has been proposed that this unconventional MR results from large spatial fluctuations in the conductivity of the material \cite{par}.
 The model used is a random resistor network, thus implying an inhomogeneous conductor.
We think that this can be the case here for at least two reasons: the first one is the grain boundaries and/or the composition
 fluctuations that can not be excluded
 in our ceramic, the other one is the proximity of the first order transition where we have measured large resistance fluctuations.
 We think that this latter proposition is the most likely, but only comparison with single crystal data could allow to be more conclusive.

\section{Magnetic properties: frequency response and non linear susceptibility}

Figure 9 presents the magnetization $M$ versus temperature measured under 0.3 $T$, both in zero field cooled and field cooled modes.
 Two major transitions are observed, as previously reported \cite{aka}. After cooling the samples from $T =$ 400 $K$, a sharp increase of $M$ is observed at $T \approx  T_1\simeq $ 300 $K$.
  M goes through a maximum and then decreases. At lower temperature, the ZFC and FC curves separate at $T\simeq $ 225 $K$. The characteristic temperatures are similar
 to the values reported by \textit{Akahoshi et al} \cite{aka}.
 We will focus here on the high $T$ transitions. 
If one analyses the data for 300 $K < T <$  400 $K$ with a Curie-Weiss law, an effective moment of $\mu_{eff} =$ 2.98 $\mu_B /Co$
 and $\theta_p =$ -14 $K$ are found. In $TbBaCo_2O_{5+x}$ or in $NdBaCo_2O_{5+x}$, a second linear part is observed below $T_1$,
 which is usually associated to a spin state transition \cite{morito}. Because of the magnetic rare earth cation,
 it is not completely clear if this change of slope should be attributed 
to the spin state of Co ions only. Our sample does not have the magnetic rare earth ion, which should remove this difficulty.
 We have studied in details the temperature dependence of the susceptibility
 close to $T_1$ using detailed ac measurements.
 We observed the same kind of slope breaking in $\chi ^{'-1}$ (see fig.10). 
Nevertheless, using linear susceptibility and in a very restricted range of temperature (a few Kelvins),
 it is difficult to be convinced that the change in the temperature dependence of the susceptibility proves a spin state transition. The first constraint is to be sure that the inflexion point
 should separate two pure paramagnetic regimes.
 Nonlinear ac susceptibility is a very sensitive tool which allow to distinguish between a pure paramagnetic regime and a magnetic ordered state, even if this latter is incomplete.
In the low field limit, the
magnetization can be expanded in power of the magnetic field $H$, which gives for ac measurements under a small ac field \cite{levy}:

\begin{equation}
\chi (\omega^{'}) =  \chi_0 + 2 \chi_1.H + 3 \chi_2.H^2 + 4
\chi_3.H^3+.....
\end{equation}

The $\chi_1$ parameter breaks the symmetry by field inversion and thus implies the existence of a spontaneous moment,
 and is uncompatible with a genuine paramagnetic state.
In our sample, we observe that $\chi_1$ becomes non zero, i.e. increases above the experimental resolution, for $T\leq$ 292 $K$ (fig.11). This shows that the slope
 breaking observed in fig.9 is not in a pure paramagnetic
 state but is certainly due to the appearance of magnetic order, not necessarily linked with a spin state transition.
 Extracting an effective moment $\mu_{eff}$ from a Curie-Weiss law in this regime in this regime is meaningless.

For $T < T_1$, both ferrimagnetic or canted antiferromagnetic states have been proposed.
 Interestingly, we observe a susceptibility peak at a temperature slightly lower than $T_1$ and
that this peak collapses when moderate dc fields ($B \geq$ 0.1 $T$) are applied.
 Furthermore, the peak is observed on heating the sample but is strongly attenuated on cooling (fig.12).
These phenomena are reminiscent of the Hopkinson effect which exists in the ferromagnetic compounds.
In the classical Stoner-Wohlfarth theory, the magnetization $M$ depends on the saturation magnetization $M_s$
 and inversely on the magnetic anisotropy $K$  \cite{stoner}. Because of the competition between their temperature dependencies, 
 a peak in the magnetization can occur close to $T_c$. It is called the Hopkinson maximum \cite{williams}. These effect
 is classically observed for ferromagnetic or ferrimagnetic to paramagnetic transition as the temperature increases (it is strongly attenatued for the reciprocal transition), just
 below the Curie temperature. Note that it can be also observed below a Neel temperature for equivalent reasons \cite{neelT}.

If a dc field is applied, a smaller secondary peak can be observed. This peak decreases in amplitude and shifts
 at higher temperature when the applied field increases. While the first (Hopkinson) peak is not critical,
the field dependencies of the secondary maximum reveal critical
fluctuations which are characteristics of a second order transition \cite{second}. The behavior of this peak is
 compatible with conventional treatment using static scaling law.
The dependence of the maximum amplitude varies as $\chi (T_{max}, B) \propto B^{1/\delta -1}$ and
 the peak temperature varies as $T_{max} - T_c \propto B^{(\gamma + \beta )^{-1}}$. We find exponents $\delta=$ 5.30 $\pm$ 0.02,
 and  $\gamma + \beta =$ 1.75 $\pm$ 0.02, not far from the Heisenberg model ($\delta=$ 4.78, $\gamma + \beta =$ 1.765) (see fig.13) \cite{heisen}.
 Moreover, except close to the Hopkinson peak (between 285 and 287.5 $K$), we have measured that the susceptibility is thermally
 reversible around $T_1$, as expected for a continuous transition.
 It is important to note that this is very different from the strong thermal hysteresis (more than 10 $K$)
 observed in the transport properties and explained
 above by the first order character of the resistive transition.

 At a lowest temperature, for $250 K \leq  T \leq T_2$, the magnetic susceptibility shows a thermal hysteresis.
 To say more on this point, we have studied the frequency dependence of the susceptibility. The ac response evidences
also strong frequency dependence for $T \leq T_2$. More precisely, a non Debye relaxation with a logarithm
 dependence of the susceptibility is observed at low frequencies (not shown).
This dependence is characteristic of interface pinning in random media \cite{natterman}, as in the case of magnetic domains in the presence of quenched disorder.
 The rather strong frequency dependence observed here indicates that the domain walls dynamic has low cooperativity, and is not compatible with glassy  .
 Finally, the non-equilibrium magnetic features are certainly coming from
 domain walls dynamics in the presence of pinning centers.
 What is important here is that the magnetic transition at $T_1$ has the characteristics of a paramagnetic-"ferromagnetic" second order transition,
 and thus can not be directly linked to the first order transition which is observed with the transport measurements.
Note that the ferromagnetic chararacter, which we observed here, arises from the observation of a spontaneous moment, of the Hopkinson peak, of the continuous nature of the transition, and of domains walls dynamics.
 Strickly speaking, all these features are also compatible with an uncompensate antiferromagnetic state, such as the one proposed very recently using neutrons diffraction results and group theory analysis \cite{group}.
 
 To conclude, the electronic transport properties of $YBaCo_2O_{5.45}$ exhibit the characteristics of a first order transition: hysteresis, relaxation, and large fluctuations.
Despite the strong relaxation effects, an equilibrium positive magnetoresistance is evidenced close to this transition, and has been attributed to spatial fluctuations of the conductivity.
 We observe the change of slope in the temperature dependence of the susceptibility, as observed in others 112 cobaltites with magnetic rare earth. Nevertheless,
 we show that it corresponds
 to the appearance of a spontaneous moment, and thus it can not be taken as a support for a spin state transition. 
AC susceptibility evidences a continuous transition and domains wall dynamics. The important conclusion is that transport and magnetic properties appear not coupled at a macroscopic scale as far as the high temperature transition is involved.

 Acknowledgements:
 
 A.P thanks Sandrine Rey-Froissard for the iodometric titration.

\newpage

\begin{figure}[htbp]
\center\includegraphics[width=6cm]{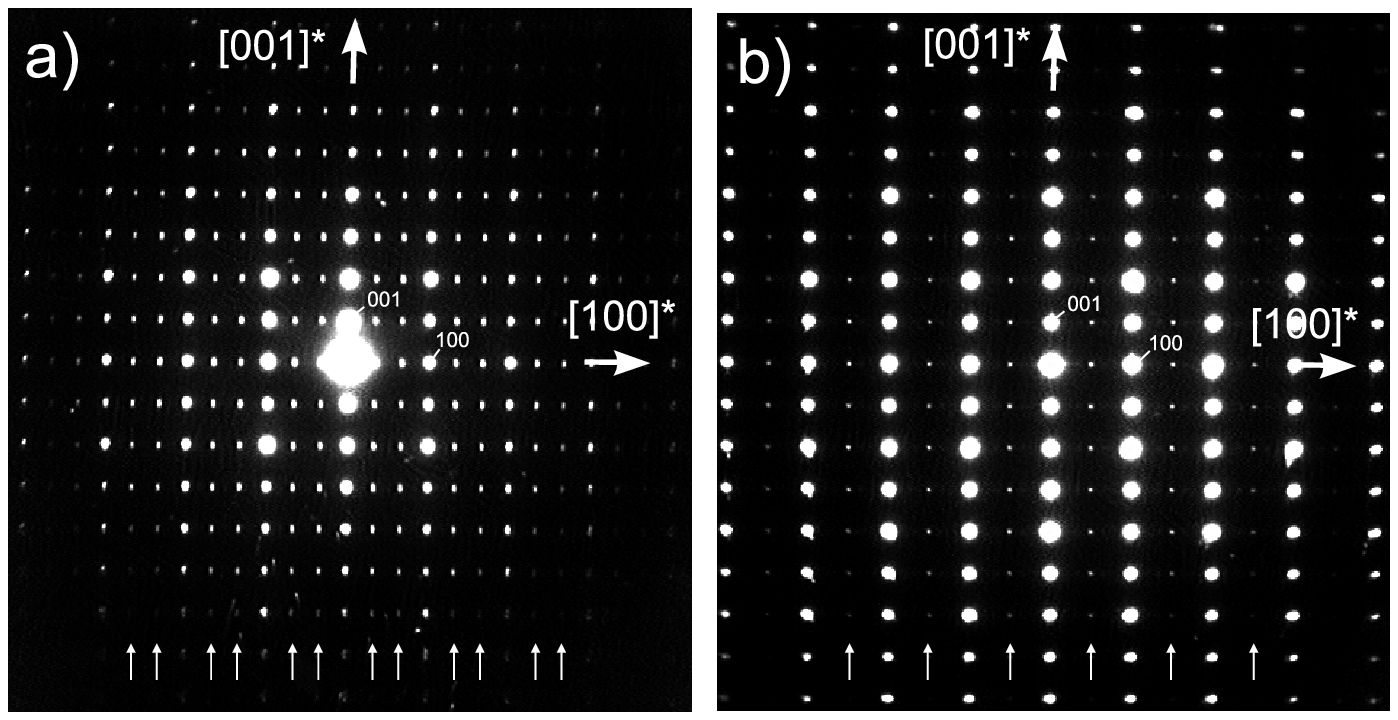}
\vspace{-0.5cm}
\caption{\footnotesize [010] zone axis patterns SAED observed for the compound YBaCo$_2$O$_{5.45\pm0.03}$ where both $3a_p$ x $3a_p$ x $2a_p$ (a) and $2a_p$ x $a_p$ x $2a_p$ (b) superstructures can be observed. The indexation is given considering the
 112 subcell ($a_p$ x $a_p$ x $2a_p$). Rows of supercell reflections are indicated by arrows. } 
\end{figure}

\vskip 2 cm

\begin{figure}[htbp]
\center\includegraphics[width=6cm]{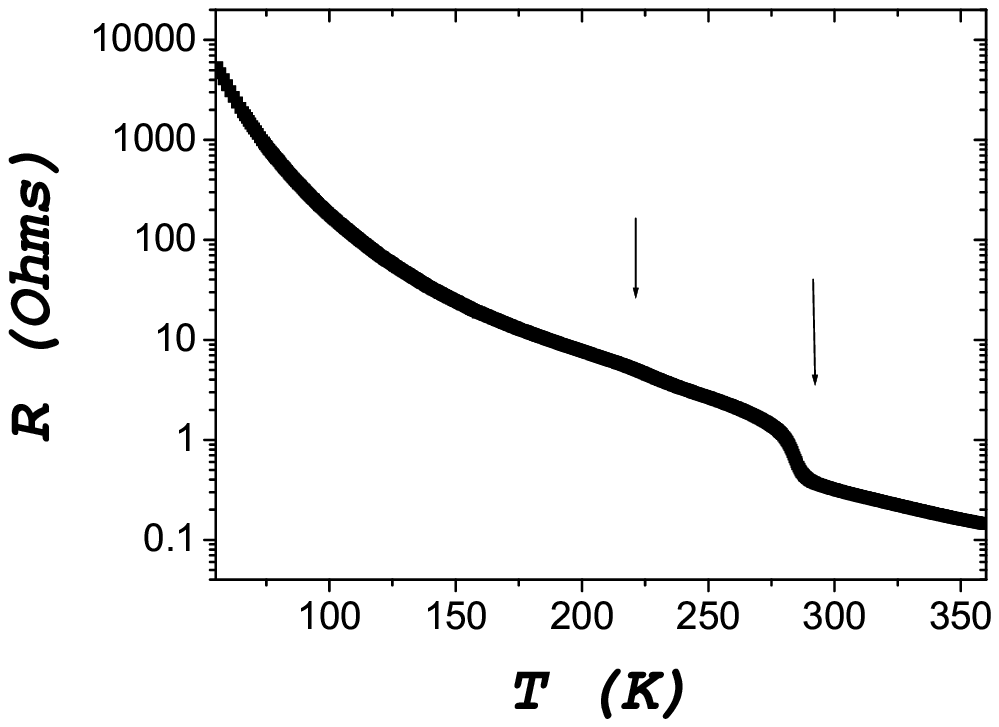}
\vspace{-0.5cm}
\center\includegraphics[width=6cm]{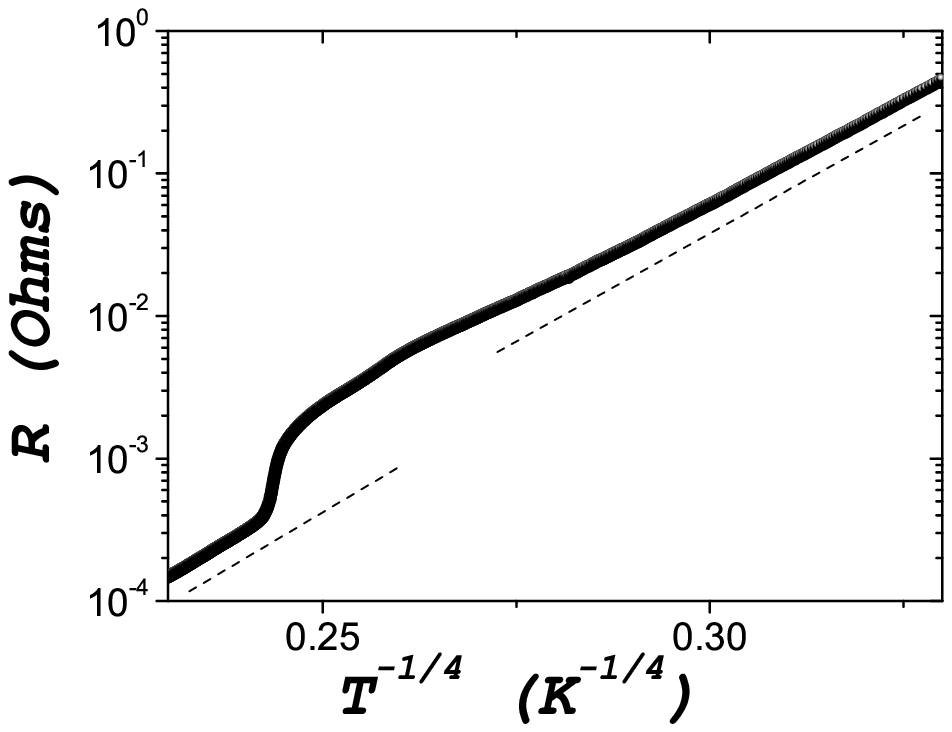}
\vspace{-0.5cm}
\caption{a/ Resistance $R$ as a function of $T$ in a semi-log scale, measured on heating the sample ($I=$ 100 $\mu  A$). The first arrow at
 290 $K$ marks the "M-I" transition. The second arrow shows a crossover also observed in magnetic susceptibility. b/ $R$ as function of $T^{-1/4}$ in a semi-log plot. 
The two dotted lines are guide for the eyes. Note that the high temperature regime is not "metallic".}
\end{figure}

\vskip 2 cm
\begin{figure}[htbp]
\includegraphics[width=5cm]{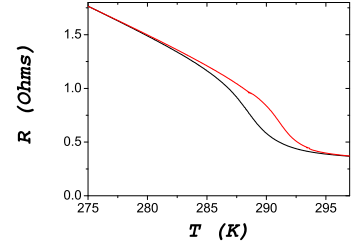}
\vskip 2 cm
\caption{Thermal hysteresis of the resistance at the "M-I" transition (cooling/heating rate= 0.1K/min).}
\end{figure}

\vskip 2 cm
\begin{figure}[htbp]
\includegraphics[width=5cm]{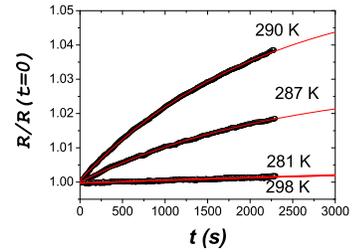}
\caption{Time dependance of the resistance after cooling the sample, for different temperatures. The solid line is
 a fit using a molecular growth model.}
\end{figure}

\vskip 2 cm
\begin{figure}[htbp]
\includegraphics[width=5cm]{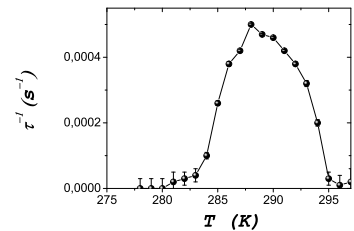}
\caption{$\tau^{-1}$ as function of the temperature. This evidences a metastable domain (a spinodal domain for the supercooled state) around the first order transition.}
\end{figure}

\vskip 2 cm
\begin{figure}[htbp]
\includegraphics[width=5cm]{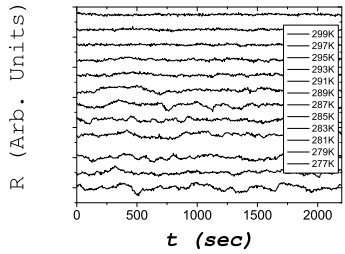}
\caption{Time series of resistances ($I=$ 100 $\mu A$, cooling from 300 $K$). For each trace, the main relaxation has been substracted, so as to reveal the resistance noise only.}
\end{figure}

\vskip 2 cm
\begin{figure}[htbp]
\includegraphics[width=5cm]{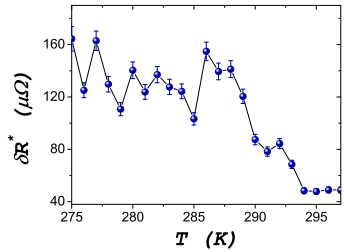}
\caption{Resistance noise as a function of the temperature ($I=$ 100 $\mu A$, cooling from 300 $K$).
 $\delta R^*$ is the noise integrated over a 0.05-5 $Hz$ bandwidth.}
\end{figure}

\vskip 2 cm
\begin{figure}[htbp]
\includegraphics[width=5cm]{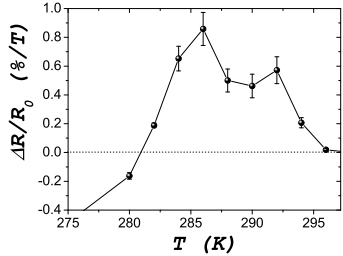}
\caption{magnetoresistance ratio ($(R(1T)-R_0)/R(0)$) as a function of the temperature. Note that the relaxation effect is also present when cycling the field but its magnitude corresponds to the error bars.}
\end{figure}

\vskip 2 cm
\begin{figure}[htbp]
\center\includegraphics[width=5cm]{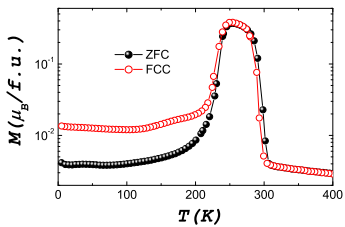}
\vskip 1 cm
\caption{FC and ZFC magnetisation as function of the temperature in a semi-log scale($B=$ 0.3T).}
\end{figure}

\vskip 2 cm
\begin{figure}[htbp]
\includegraphics[width=5cm]{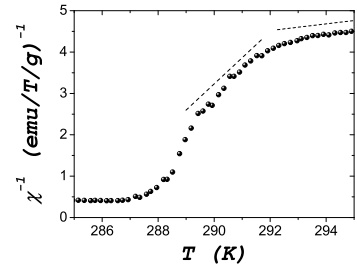}
\caption{ $\chi'^{-1}$ as function of the temperature. Note the change of slope at $T \approx $ 292 $K$.}
\end{figure}

\vskip 2 cm
\begin{figure}[htbp]
\includegraphics[width=5cm]{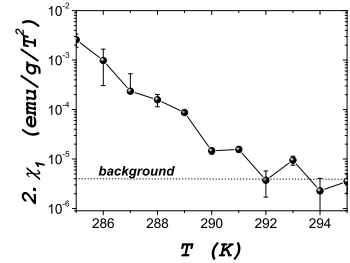}
\caption{ $\chi'_{1}$ as function of the temperature. This shows that a spontaneous moment appears for $T \leq$  292 $K$.}
\end{figure}

\vskip 2 cm
\begin{figure}[htbp]
\includegraphics[width=5cm]{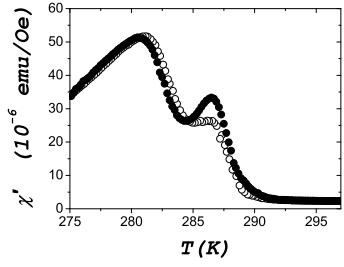}
\caption{Zero field limit of the in-phase ac susceptibility. Note the peak observed only on heating the sample (plain points), and associated to the Hopkinson maximum.}
\end{figure}

\vskip 2 cm
\begin{figure}[htbp]
\includegraphics[width=5cm]{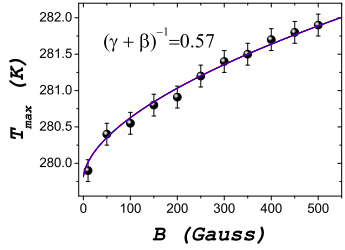}
\vskip 1 cm
\includegraphics[width=5cm]{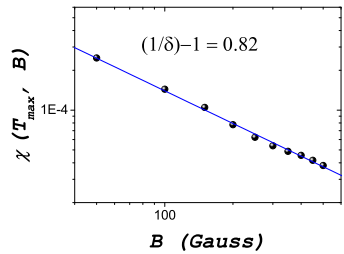}
\caption{Scaling of the susceptibility peak and of its temperature as function of the magnetic field. The solid lines are power law fits according
 to the static scaling laws of second order transition (see text).}
\end{figure}

\end{document}